\documentclass[prb,twocolumn,superscriptaddress,showpacs,preprintnumbers,amsmath,amssymb]{revtex4}

\usepackage{graphicx}
\usepackage{dcolumn}
\usepackage{bm}
\usepackage{ifpdf}
\usepackage{multirow}

\begin{document}

\bibliographystyle{apsrev} 


\title{Accelerating an adiabatic process by nonlinear sweeping}

\author{Xingxin Cao}
\affiliation{Department of Optical Science and Engineering, Fudan University, Shanghai 200433, China}

\author{Jun Zhuang}
\affiliation{Department of Optical Science and Engineering, Fudan University, Shanghai 200433, China}

\author{X.-J. Ning}
\affiliation{Institute of Modern Physics, Department of Nuclear Science and Technology, Fudan University, Shanghai 200433, China}

\author{Wenxian Zhang}
\email{wenxianzhang@fudan.edu.cn}
\affiliation{Department of Optical Science and Engineering, Fudan University, Shanghai 200433, China}

\date{\today}

\begin{abstract}
We investigate the acceleration of an adiabatic process with the same survival probability of the ground state by sweeping a parameter nonlinearly, fast in the wide gap region and slow in the narrow gap region, as contrast to the usual linear sweeping. We find the expected acceleration in the Laudau-Zener tunneling model and in the adiabatic quantum computing model for factorizing the number $N=21$.
\end{abstract}


\maketitle

Due to its robustness against noises, adiabatic quantum computing (AQC) is one of the promising quantum computation schemes which are experimentally feasible~\cite{Xu11, Childs01}. However, the main concern is that the energy gap between the ground state and the lowest excited state of a quantum computer decreases exponentially with the increasing number of qubits~\cite{Georgeot00a, Georgeot00b}, so that the adiabatic evolution time required by the algorithm would increase exponentially and eventually cancels the speedup benefit of a quantum computer.

One way to overcome the long-adiabatic-evolution-time obstacle is using rapid adiabatic passage method~\cite{Allen75, Chen10, Chen10a, Li11}, which is capable of transferring an initial ground state to a final ground state within an arbitrary short time. The price is that all the gap information must be provided, which is usually beyond the reach of current techniques for a many-qubit quantum computer. It is thus useful to accelerate an adiabatic process, similar to the rapid adiabatic passage method, but with less information about the gap, for instance, only the extremes of the gap during the total evolution.

In this paper, we investigate nonlinear sweeping methods to accelerate adiabatic processes, including the famous Laudau-Zener (LZ) tunneling model and the AQC model~\cite{Wittig05, Peng08}, with only the knowledge of the extremes of the gap. For the LZ model, we numerically confirm that the acceleration effect of a nonlinear sweeping. We then develop a similar nonlinear sweeping scheme for a simplified AQC model and finally confirm the effectiveness of this nonlinear sweeping method by factorizing the number $N=21$.


\begin{figure}
  \includegraphics[width=3.5in]{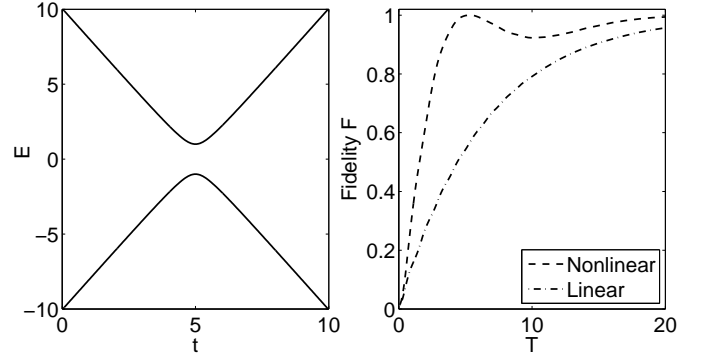}
  \caption{The instantaneous energy level diagram for a two-level LZ model (left) and the fidelity of the final ground state (right). The parameters are $\omega_0 = 10$, $\omega_x = 1$, and $T=10$ for the left panel. We set $\hbar = 1$ for convenience. In the right panel, the dash-dotted line denotes the fidelity under the linear sweeping and the dashed line the nonlinear sweeping. The fidelity for the nonlinear sweeping is higher than that for the linear sweeping.}
  \label{fig:LZ}
\end{figure}

The LZ model of a two-level system is described by
\begin{equation}\label{eq:LZ}
    H(t) =  \hbar \omega_x \sigma_x + \hbar \omega_z(t) \sigma_z,
\end{equation}
where $\omega_{x,z}$ is the frequency of the Zeeman splitting along the $x$ or $z$ direction, and $\sigma_{x,z}$ the Pauli matrix. For a symmetric system as shown in the left panel of Fig.~\ref{fig:LZ}, the time dependence of a linear sweeping of $\omega_z(t)$ is
\begin{equation}\label{eq:linear}
    \omega_z(t) = -\omega_0 + v t,
\end{equation}
where $v=2\omega_0/T$ denotes the sweeping velocity, with $T$ being the total evolution time. If $\omega_z(t)$ sweeps fast from $-\omega_0$ to $\omega_0$, an initial ground state may have a large probability to be transferred to the final excited state. Such a probability is described by the Laudau-Zener formula~\cite{Wittig05}. If $\omega_z(t)$ slowly changes, the system would stay in the ground state.

It is for sure that the linear sweeping of $\omega_z(t)$ is not optimized for the fidelity $F$, which measures the survival probability of the final ground state and is defined as
\begin{equation}\label{eq:fid}
    F = |\langle \psi(T) | \psi_g \rangle|^2
\end{equation}
with $|\psi_g\rangle$ denoting the instantaneous ground state of the Hamiltonian $H=H(T)$ and $|\psi(T)\rangle$ the actually evolved wave function at the end time $t=T$. An intuition for the optimization is sweeping fast if the energy gap is large but sweeping slow if the gap is small, where the energy gap $\Delta$ is defined as the energy  difference between the ground state $\omega_g$ and the lowest excited state $\omega_e$, i.e., $\Delta = \omega_e - \omega_g$. It is straightforward to construct a nonlinear sweeping form which satisfies the above intuition,
\begin{equation}\label{eq:quad}
    \omega_z(t) = \left\{\begin{matrix}
                    -\omega_0 \left({t \over t_c}-1\right)^2, & \;t\le t_c \\
                    \;\omega_0 \left({t \over t_c}-1\right)^2, & \;t> t_c \\
                  \end{matrix}\right.
\end{equation}
with $t_c = T/2$. The gap $\Delta$ is minimal at $t_c$. The corresponding sweep velocity is
\begin{equation}\label{eq:v}
    v(t) = 2\omega_0\left|{t\over t_c} -1\right|,
\end{equation}
which is large near $t=0$ and $T$ where the gap is large but small near $t=t_c$ where the gap is small. The velocity for the whole evolution can be adjusted by changing the total time $T$. The longer the $T$ is, the smaller the velocity is.

We present in the right panel of Fig.~\ref{fig:LZ} the fidelity $F$ of the final ground state under either the linear sweeping form Eq.~(\ref{eq:linear}) or the nonlinear sweeping form Eq.~(\ref{eq:quad}). It clearly shows that the fidelity with nonlinear sweeping is much higher than that with linear sweeping, provided that the total evolution time $T$ is the same. In other words, the adiabatic LZ process can be accelerated by adopting a nonlinear sweeping form.

We note that other more complex nonlinear forms also accelerate the adiabatic process, especially the stimulated Raman adiabatic passage or the hyperbolic tangent form~\cite{Allen75,Chen10}, and the fidelity $F$ may be even higher. These methods usually require all the information of the energy levels which is not always available. While in the nonlinear formula, only the position of the minimal gap $t_c$ is required. In this sense, the nonlinear sweeping method is easier to implement in experiments.

We consider next a time-dependent two-level system, that is a single qubit AQC model,
\begin{equation}\label{eq:mid}
    H(t) = [1-s(t)] \omega_x \sigma_x + s(t) \omega_z \sigma_z,
\end{equation}
where $s(t)$ is increasing monotonically with $s(0)=0$ and $s(T)=1$. By sweeping slowly $s(t)$ from 0 to 1, the two-level system transfers from an initial ground state to its final ground state. A linear sweeping and a quadratic sweeping form used in experiment~\cite{Peng08, Xu11} are, respectively,
\begin{equation}\label{eq:ml}
    s(t) = {t\over T}
\end{equation}
and
\begin{equation}\label{eq:mq}
    s(t) = \left({t\over T}\right)^2.
\end{equation}
It is obvious that neither the linear sweeping nor the quadratic sweeping satisfies the requirement of fast sweeping in large gap region and slow sweeping in small gap region. We find the optimal sweeping form as follows.

It is the best to do the similar procedure to the above LZ model and utilize the optimized sweeping scheme. We alternate the single qubit AQC Hamiltonian Eq.~(\ref{eq:mid}) to a LZ model Eq.~(\ref{eq:LZ}) where only one of the two parameters before the Pauli matrices is time-dependent
\begin{equation}\label{eq:al}
    H(t) = \omega_\perp \sigma_\perp + \omega_n(s(t)) \sigma_n,
\end{equation}
where
\begin{eqnarray*}
  \sigma_\perp &=& \sigma_z \sin\theta + \sigma_x \cos\theta, \\
  \sigma_n &=& \sigma_z \cos\theta - \sigma_x \sin\theta, \\
  \omega_\perp &=& \omega_x \cos\theta, \\
  \omega_n(t) &=& s(t) \Omega - \omega_x \sin\theta
\end{eqnarray*}
with $\cos\theta = \omega_z / \Omega$, $\sin\theta = \omega_x / \Omega$, and $\Omega = \sqrt{\omega_z^2 + \omega_x^2}$.

To locate the position of the minimal gap at $s(t_c) = s_c$, we calculate the energy gap
\begin{equation}\label{eq:gap}
    \Delta(s) = 2\sqrt{s^2 \omega_0^2 + (1-s)^2 \omega_x^2}
\end{equation}
where the minimum satisfies
\begin{equation}\label{eq:deriv}
    \left.{d\Delta \over ds}\right|_{s_c} = 0.
\end{equation}
It is easy to find that
\begin{equation}\label{eq:sc}
    s_c = {\omega_x^2 \over \omega_x^2 + \omega_0^2}.
\end{equation}
Another way to obtain the above result of $s_c$ is setting $\omega_n(s_c) = 0$.

To utilize the optimized quadratic sweeping scheme, for the LZ model we notice that $s$ in the AQC model Eq.~(\ref{eq:al}) takes the place of $t$ in the LZ model Eq.~(\ref{eq:LZ}), so we use the following relation
\begin{equation}\label{eq:s0}
    {d\omega_n(s(t)) \over dt} = A|s(t)-s_c|.
\end{equation}
The value of $A$ can be determined by the boundary of $\omega_n(s(t))$. This equation implies that
\begin{equation}\label{eq:s}
    {ds\over dt} = {A\over \Omega} |s-s_c|,
\end{equation}
which gives the relation $s(t)$. Similar to the optimized scheme in the LZ model, the form of $s(t)$ sweeps fast in the large gap region and slow in the small gap region with respect to $s$.

\begin{figure}
  \includegraphics[width=3.5in]{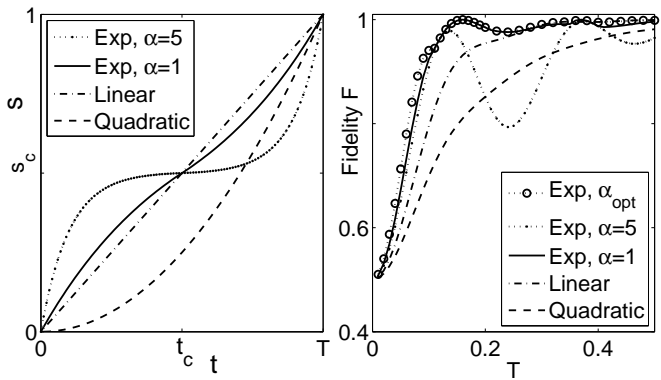}
  \caption{Left: Various forms of $s(t)$; Right: Dependence of the fidelity on the total evolution time. Dotted line with circles --- Exponential-like with opitimized $\alpha$; Dotted lines --- Exponential-like with $\alpha=5$; Solid lines --- Exponential-like with $\alpha=1$; Dash-dotted lines --- Linear; Dashed lines --- Quadratic. The parameters are $\omega_x = 18$ and $\omega_z = 30$. The Exponential-like schemes are better than the Linear of the Quadratic one.}
  \label{fig:AQC}
\end{figure}

It is straightforward to obtain the relation $s(t)$, which satisfies not only the initial and final values but also an extreme point $s(t_c) = s_c$,
\begin{equation}\label{eq:st}
    s(t) = \left\{\begin{matrix}
                    {s_c\over 1-e^{-\alpha}} (1-e^{-\alpha t/t_c}), & t\le t_c, \\
                    1-{1-s_c \over e^{\alpha/s_c}-e^\alpha} (e^{\alpha/s_c}-e^{\alpha t/t_c}), & t>t_c \\
                  \end{matrix}
                  \right..
\end{equation}
We have introduced an adjustable parameter $\alpha$ varies the curvature of $s(t)$ [see the left panel of Fig.~\ref{fig:AQC}]. When $\alpha = 0$, $s(t)$ is linear; When $\alpha = 1$, $s(t)$ is a two-interval exponential-like piecewise function; When $\alpha \rightarrow \infty$, $s(t)$ is step-like function. To determine $t_c$, we adopt a simplest form $t_c/T = s_c$.

We present in the right panel of Fig.~\ref{fig:AQC} the fidelity of the final ground state under the time-dependent Hamiltonian Eq.~(\ref{eq:mid}) for various sweeping schemes. Note that the initial state of the two-level system is an eigenstate of $\sigma_x$, which overlaps with the final instantaneous ground state so the fidelity is $0.5$ even if $T$ approaches zero. As shown in the right panel of Fig.~\ref{fig:AQC}, it is obvious that the quadratic scheme is the worst choice. The fidelity with the linear scheme is between that with quadratic scheme and that with exponential-like scheme. By optimizing $\alpha$ for each T in the exponential-like schemes, the fidelity reaches its highest value. We also notice that for some values of $\alpha$ in the exponential-like schemes, the fidelity oscillates if the total evolution time $T$ becomes large, e.g., $\alpha = 5$. These oscillations might be due to the over-long-time stay of the system near the minimal gap region where the gap is almost constant~\cite{Zhang06}.

The time-dependent Hamiltonian for factorizing $N=21$ is an interpolation of an easy-realized initial Hamiltonian $H_0$ and a problem-solved Hamiltonian $H_P$~\cite{Peng08},
\begin{equation}\label{eq:aqc}
    H(t) = [1-s(t)]\; H_0 + s(t)\; H_P,
\end{equation}
where $s(t)$ changes from $0$ to $1$ as the system evolves and
\begin{eqnarray*}
    H_0 &=& g(\sigma_{1x} + \sigma_{2x} + \sigma_{3x}), \\
    H_P &=& [N-(2I-\sigma_{1z})(4I-\sigma_{2z}-2\sigma_{3z})]^2.
\end{eqnarray*}
In the above equations, $g$ is the Zeeman splitting of a uniform magnetic field along $x$ direction, $I$ is the 2-by-2 identity matrix, and $\sigma_{i\alpha}$ with $i=1,2,3$ and $\alpha=x,z$ denotes the $\alpha$ Pauli matrix for the $i$th spin. The ground state of $H_P$, $|\downarrow_1\rangle \otimes|\downarrow_2\downarrow_3\rangle$ interpreted as $3\otimes 7$, is the solution to the factors of $N$. The 3-spin system is prepared initially in the ground state of $H_0$. After an adiabatic evolution under the total Hamiltonian $H(t)$, the system eventually reaches the ground state of $H(T)= H_P$ and the number $N$ is factorized.

\begin{figure}
  \includegraphics[width=3.5in]{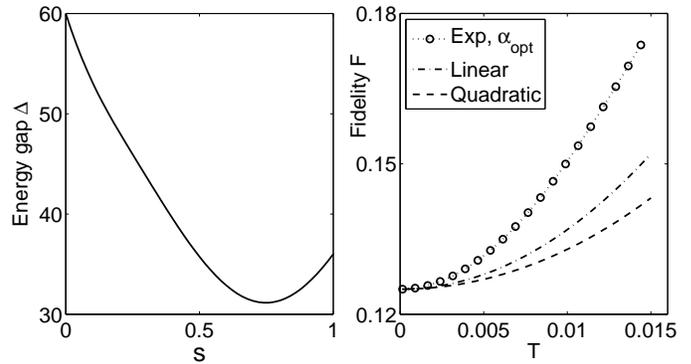}
  \caption{Left: The instantaneous energy gap versus $s$ for $N=21$. The minimal gap position is near $s_c\approx 0.74$. Right: The dependence of the fidelity on the total evolution time for factorizing $N$ with optimized exponential-like sweeping scheme (dotted line with circles), linear sweeping scheme (dash-dotted line), and quadratic sweeping scheme (dashed line). The parameter is $g = 30$. Among the three schemes, the optimized exponential-like one shows the highest fidelity at the same evolution time $T$ and is the best choice.}
  \label{fig:N21}
\end{figure}

Usually, two sweeping schemes are experimentally adopted~\cite{Peng08, Xu11}. One is the linear scheme as defined in Eq.~(\ref{eq:ml}) and the other is the quadratic scheme as in Eq.~(\ref{eq:mq}). We have shown in the two-level model that both the linear and the quadratic sweeping schemes are not the optimal. We may use the exponential-like scheme to accelerate the AQC process if the energy gap between the ground state and the lowest excited state is approximately in an inverted parabolic shape with respect to $s$. Fortunately, the gap for $N=21$ is indeed very similar to that for the single qubit AQC model system, as shown in the left panel of Fig.~\ref{fig:N21}. Thus we directly apply the exponential-like sweeping scheme to the adiabatic process $N=21$ factorization.

We present in the right panel of Fig.~\ref{fig:N21} the fidelities of the evolved final state to the ground state of $H_P$ at different total evolution time $T$ with the linear sweeping scheme, the quadratic scheme, and the exponential-like scheme. We limit ourselves in the short $T$ region where the difference among these schemes are prominent. It is clearly shown in the figure that the fidelity with the exponential-like sweeping scheme is higher than that with either the linear or the quadratic sweeping scheme for the same total evolution time $T$. Thus, to reach the same final fidelity, the exponential-like sweeping scheme uses the shortest time and accelerates the AQC process.

For larger $N$, more qubits are required to factorize the number. The energy levels inevitably become so complex that the energy gap $\Delta$ may show multiple extremes. In this situation, a more complecated piecewise function $s(t)$ with exponential-like form in each interval might be more appropriate. More explorations along this direction is worthy in the future.

In conclusion, we develop a nonlinear sweeping scheme for an adiabatic process, which changes fast in the large gap region and slow in the small gap region. This scheme improves the survival probability of the final ground state and may thus accelerate the adiabatic process. We confirm the usefulness of this nonlinear scheme in the Laudau-Zener tunneling model, the single qubit AQC model, and the AQC of factorizing the number $N=21$. The proposed nonlinear scheme may be useful in AQC experiments to shorten the adiabatic evolution time.

XC and WZ acknowledge support by the National Natural Science Foundation of China under Grant No. 10904017, NCET, Specialized Research Fund for the Doctoral Program of Higher Education of China under Grant No. 20090071120013, and Shanghai Pujiang Program under Grant No. 10PJ1401300.



\end{document}